\documentclass[preprint]{aastex}

\usepackage{latexsym}		
\usepackage{graphicx}		
\usepackage{rotating}		
\usepackage{natbib}  
\usepackage{savesym}
\usepackage{amssymb}
\savesymbol{doublespace}
\usepackage{xspace}
\usepackage{color}
\usepackage{multicol}
\usepackage{mdframed}
\usepackage{url}
\usepackage{subfigure}
\usepackage{lscape}
\usepackage{grffile}
\usepackage{standalone}
\standalonetrue
\usepackage{import}
\usepackage[utf8]{inputenc}
\usepackage{longtable}
\usepackage{booktabs}

\newcommand{\rhoPDF}{\ensuremath{\rho-\mathrm{PDF}}\xspace}
\newcommand{\meanrho}{\ensuremath{\langle\rho\rangle}\xspace}
\newcommand{\GRSMC}{GRSMC 43.30-0.33\xspace}
\newcommand{\north}{G43.17+0.01\xspace}
\newcommand{\south}{G43.16-0.03\xspace}

\newcommand{\msun}{\ensuremath{M_{\odot}}\xspace}			

\newcommand{\hh}{\ensuremath{\textrm{H}_{2}}\xspace}			
\newcommand{\formaldehyde}{\ensuremath{\textrm{H}_2\textrm{CO}}\xspace}

\newcommand{\ortho}{\ensuremath{\textrm{o-H}_2\textrm{CO}}\xspace}
\newcommand{\oneone}{\ensuremath{1_{10}-1_{11}}\xspace}
\newcommand{\twotwo}{\ensuremath{2_{11}-2_{12}}\xspace}

\newcommand{\kms}{\textrm{km~s}\ensuremath{^{-1}}\xspace}	

\newcommand{\percc}{\ensuremath{\textrm{cm}^{-3}}\xspace}
\newcommand{\persc}{\ensuremath{\textrm{cm}^{-2}}\xspace}

\newcommand{\perpc}{\textrm{pc}\ensuremath{^{-1}}\xspace}
\newcommand{\perkms}{\textrm{per~km~s}\ensuremath{^{-1}}\xspace}	
\newcommand{\perkmspc}{\perkms\perpc}	
\newcommand{\um}{\ensuremath{\mu \textrm{m}}\xspace}    

\newcommand{\twelveco}{\ensuremath{^{12}\textrm{CO}}\xspace}
\newcommand{\thirteenco}{\ensuremath{^{13}\textrm{CO}}\xspace}
\newcommand{\ceighteeno}{\ensuremath{\textrm{C}^{18}\textrm{O}}\xspace}
\def\ee#1{\ensuremath{\times10^{#1}}}
\newcommand{\degrees}{\ensuremath{^{\circ}}}

\def\eqref#1{Equation \ref{#1}}

\def\Figure#1#2#3#4#5{
\begin{figure*}[htp]
\includegraphics[scale=#4,angle=#5]{#1}
\caption{#2}
\label{#3}
\end{figure*}
}

\def\FigureTwo#1#2#3#4#5{
\begin{figure*}[htp]
\epsscale{#5}
\plottwo{#1}{#2}
\caption{#3}
\label{#4}
\end{figure*}
}

\def\FigureFourPDF#1#2#3#4#5#6#7{
\begin{figure*}[htp]
\subfigure[]{ \includegraphics[width=#7]{#1} }
\subfigure[]{ \includegraphics[width=#7]{#2} }
\subfigure[]{ \includegraphics[width=#7]{#3} }
\subfigure[]{ \includegraphics[width=#7]{#4} }
\caption{#5}
\label{#6}
\end{figure*}
}

\def\Table#1#2#3#4#5#6{
\begin{deluxetable}{#1}
\tablewidth{0pt}
\tabletypesize{\footnotesize}
\tablecaption{#2}
\tablehead{#3}
\startdata
\label{#4}
#5
\enddata
\bigskip
#6
\end{deluxetable}
}

\begin{document}
\ifstandalone
\newcommand{\casa}{1}
\newcommand{\eso}{2}
\newcommand{\monash}{3}

\author{Adam Ginsburg\altaffilmark{\casa,\eso},
        Christoph Federrath\altaffilmark{\monash},
        Jeremy Darling\altaffilmark{\casa}}
\email{Adam.G.Ginsburg@gmail.com}

\affil{{$^\casa$}{\it{CASA, University of Colorado, 389-UCB, Boulder, CO 80309}}}
\affil{{$^\eso$}{\it{European Southern Observatory, Karl-Schwarzschild-Strasse 2, D-85748 Garching bei München, Germany}}}
\affil{{$^\monash$}{\it{Monash Centre for Astrophysics, School of Mathematical Sciences, Monash University, Vic 3800, Australia}}}

\title{
A measurement of the turbulence-driven density distribution in a
non-star-forming molecular cloud
}
\begin{abstract}
    Molecular clouds are supersonically turbulent.  This turbulence governs
    the initial mass function and the star formation rate.  In order to
    understand the details of star formation, it is therefore essential to understand
    the properties of turbulence, in particular the probability distribution of
    density in turbulent clouds.
    We present \formaldehyde volume density measurements of a non-star-forming
    cloud along the line of sight towards W49A. We use these measurements in
    conjunction with total mass estimates from \thirteenco to infer the shape of the
    density probability distribution function.  This method is complementary to
    measurements of turbulence via the column density distribution and should
    be applicable to any molecular cloud with detected CO.  We show that
    turbulence in this cloud is probably compressively driven, with a
    compressive-to-total Mach number ratio $b = \mathcal{M}_C/\mathcal{M}>0.4$.  
    We measure the standard deviation of the density distribution, 
    constraining it to the range $1.5 < \sigma_s < 1.9$
    assuming that the density is lognormally distributed.  This measurement
    represents an essential input into star formation laws.
    The method of averaging over different excitation conditions to produce a 
    model of emission from a turbulent cloud is generally applicable to optically
    thin line observations.



\end{abstract}
\fi

\section{Introduction}
Nearly all gas in the interstellar medium is supersonically turbulent.  The
properties of this turbulence, most importantly the shape of the density
probability distribution function (\rhoPDF), are essential for determining how star
formation progresses.
There are now predictive theories of star formation that include formulations
of the Initial Mass Function \citep[IMF;][]{Padoan2002a, Padoan2007a, Hennebelle2008a, Hennebelle2009a, Chabrier2010a, 
Elmegreen2011a, Hopkins2012b, Hennebelle2013a} and the star
formation rate
\citep[SFR;][]{Krumholz2005c,  Hennebelle2011a,
Padoan2011b, Krumholz2012b, Federrath2012a, Padoan2012a, Federrath2013a}.
The distribution of stellar masses and the overall star formation rate depend
critically on the \rhoPDF established by turbulence.  It is therefore essential to
measure the \rhoPDF in the molecular clouds that produce stars.

Recent works have used simulations to characterize the density distribution
from different driving modes of turbulence
\citep{Federrath2008a,Federrath2009a,Federrath2010a,Federrath2011a,Price2011b,Federrath2013a}.
These studies determined that there is a relation between the mode of turbulent driving and the width
of the lognormal density distribution, with the lognormal width (variance) $\sigma_{s}^2 = \ln\left(1+b^2
\mathcal{M}^2 \frac{\beta}{\beta+1}\right)$, where $\beta=2 (\mathcal{M}_A /
\mathcal{M})^2 = 2 (c_s/v_A)^2$ with sound speed $c_s$ and Alfven speed $v_A$,
and the logarithmic density contrast $s\equiv\ln(\rho/\rho_0)$ 
\citep[][]{Padoan2011b,Molina2012a}.

The parameter $b$ describes the coupling between the density contrast and the
Mach number \citep{Federrath2008a,Federrath2010a}.  A conceptual justification
for the parameter is that for solenoidal (curly) driving, only 1 of the 3
available spatial directions is directly compressed (longitudinal waves) and
thus $b=1/3$. Under compressive (convergent or divergent) driving, the gas is
compressed in all three spatial directions, which gives $b=3/3=1$.
\citet{Federrath2008a} and \citet{Federrath2010a} showed that simulations
driven with these modes achieve $b$ values consistent with this interpretation.

All of the above turbulence-based theories of star formation explicitly assume a
lognormal form for the density probability distribution $P_V(s)$ of the gas. 
However, recent simulations \citep{Kritsuk2007a, Schmidt2009a, Federrath2010a, Konstandin2012a, Federrath2013a, Federrath2013b} and theoretical work
\citep{Hopkins2013a} have shown that the assumption of a
lognormal distribution is often very poor; theoretical intermittent distributions and 
simulated \rhoPDF s deviate from lognormal by orders of magnitude
at the extreme ends of the density distributions.  Since these theories all involve
an integral over the density probability distribution function (PDF), deviation from
the lognormal distribution can drastically affect the overall predicted star formation
rate \citep[e.g.][]{Cho2011a,Collins2012a} and initial mass function.
Note that the modifications to the \rhoPDF driven by gravitational collapse
are unlikely to change the SFR or the IMF since gravitational overdensities have already
separated from the turbulent flow that created them
\citep{Klessen2000a,Kritsuk2011a,Federrath2012a,Federrath2013a}. It is therefore crucial that
studies of turbulence focus on clouds that are not yet dominated by
gravitational collapse (such as the cloud selected for this study) in order to study the initial conditions of star formation.

While simulations are powerful probes of wide ranges of parameter space, no
simulation to date is capable of including all of the physical processes and spatial
scales relevant to turbulence and star formation.  Observations are required to provide additional
constraints on properties of interstellar turbulence and guide simulators
toward the most useful conditions and processes to include.
\citet{Brunt2010c}, \citet{Kainulainen2012a} and \citet{Kainulainen2013a}
provide some of the first observational constraints on the mode of turbulent
driving using extinction-derived column density distributions.
They measure the parameter $b\approx0.4-0.5$, indicating that there is a `natural' mix of solenoidal
and compressive modes.  A `natural' mixture (a 2:1 mixture) of solenoidal and
compressive modes injected by the turbulent driver, i.e., a forcing ratio
$F_{comp}/F_{sol} = 1/2$, yields $b\sim0.4$. Thus, $b>0.4$ implies an enhanced
compressive forcing component relative to the naturally mixed case \citep[see
Figure 8 in][]{Federrath2010a}.

Formaldehyde, \formaldehyde, is a unique probe of density in molecular clouds
\citep{Mangum1993b}.
Like CO, it is ubiquitous, with a nearly constant abundance wherever CO is
found \citep{Mangum1993a,Tang2013a}.  The lowest excitation transitions of
\ortho at 2 and 6 cm can be observed in absorption against the cosmic microwave
background or any bright continuum source \citep{Ginsburg2011a,Darling2012b}.
The ratio of these lines is strongly sensitive to the local density of \hh, but
it is relatively insensitive to the local gas temperature
\citep{Troscompt2009a,Wiesenfeld2013a}.  The \formaldehyde line ratio has a
direct dependence on the density that is nearly independent of the column
density.  This feature is unlike typical methods of molecular-line based density
inference in which the density is inferred to be greater than the critical density
of the detected transition.

However, the particular property of the \formaldehyde densitometer we exploit
here is its ability to trace the \emph{mass-weighted} density of the gas.
Typical density measurements from \thirteenco or dust measure the total mass
and assume a line-of-sight geometry, measuring a \emph{volume-weighted}
density, i.e. $\meanrho_V = M_{tot}/V_{tot}$.  In contrast, the \formaldehyde
densitometer is sensitive to the density at which most mass resides;
this fact will be demonstrated in greater detail in Section \ref{sec:turbulenceh2co}.
The volume- and mass- weighted densities have different dependencies on the
underlying density distributions, so in clouds dominated by turbulence, if we
have measurements of both, we can constrain the shape of the \rhoPDF and
potentially the driving mode.

In \citet{Ginsburg2011a}, we noted that the \formaldehyde densitometer revealed
\hh densities much higher than expected given the cloud-average densities from
\thirteenco observations.  The densities were too high to be explained by a
lognormal density distribution consistent with that seen in local clouds.
However, this argument was made on the basis of a statistical comparison of
``cloud-average'' versus \formaldehyde-derived density measurements and left open
the possibility that we had selected especially dense clouds.  In this paper,
we use the example of a single cloud to demonstrate that the high \formaldehyde
densities must be caused by the shape of the density distribution and to infer the
shape of this distribution.

Section \ref{sec:observations} is a discussion of the observations used and the
cloud selected  for this study.  Section  \ref{sec:modeling} discusses the
\formaldehyde line and the tools used to model it.  Section
\ref{sec:turbulenceh2co} discusses the effect of turbulence on the \formaldehyde
lines and the constraints our observations place on the gas density
distribution.

\section{Observations}
\label{sec:observations}
We report \formaldehyde observations performed at the Arecibo Radio
Observatory\footnote{The Arecibo Observatory is operated by SRI International
under a cooperative agreement with the National Science Foundation
(AST-1100968), and in alliance with Ana G. Méndez-Universidad Metropolitana,
and the Universities Space Research Association.} and the Robert C. Byrd Green
Bank Telescope (GBT)\footnote{The National Radio Astronomy Observatory is a
facility of the National Science Foundation operated under cooperative
agreement by Associated Universities, Inc.} that
have been described in more detail in \citet{Ginsburg2011a}, with additional
data to be published in a future work.  The GBT observations were done in program
GBT10B/019 and the Arecibo observations as part of project a2584.  Arecibo and
the GBT have FWHM $\approx50$\arcsec\ beams at the observed frequencies of
4.829 and 14.488 GHz respectively.  Observations were carried out in a single
pointing position-switched mode with 3 and 5.5\arcmin\ offsets for the Arecibo
and GBT observations respectively; no absorption was found in the off position
of the observations described here.  The data were taken at 0.25 \kms
resolution with 150 second on-source integrations for both lines.  The continuum
calibration uncertainty is $\sim 10\%$.

The Boston University / Five-College Radio Astronomy Observatory Galactic Ring
Survey (GRS) \thirteenco data was also used.  The GRS \citep{Jackson2006a}
is a survey of the Galactic plane in the \thirteenco\ 1-0 line with $\sim
46\arcsec$ resolution.  We used reduced data cubes of the $\ell=43$ region.

\subsection{\GRSMC A non-star-forming molecular cloud}

We examine the line of sight toward \north, also known as W49A.  In a large
survey, we observed two lines of sight toward W49, the second at \south.  Both
are very bright radio continuum sources, and two foreground GMCs are easily
detected in both \formaldehyde absorption and \thirteenco emission.  Figure
\ref{fig:w49fullspec} shows the spectrum dominated by W49 itself, but with
clear \formaldehyde foreground absorption components.  The continuum levels
subtracted from the spectra are 73 K at 6 cm and 11 K at 2 cm for the south
component (\south), and 194 K at 6 cm and 28 K at 2 cm for the north component
(\north).

\FigureTwo{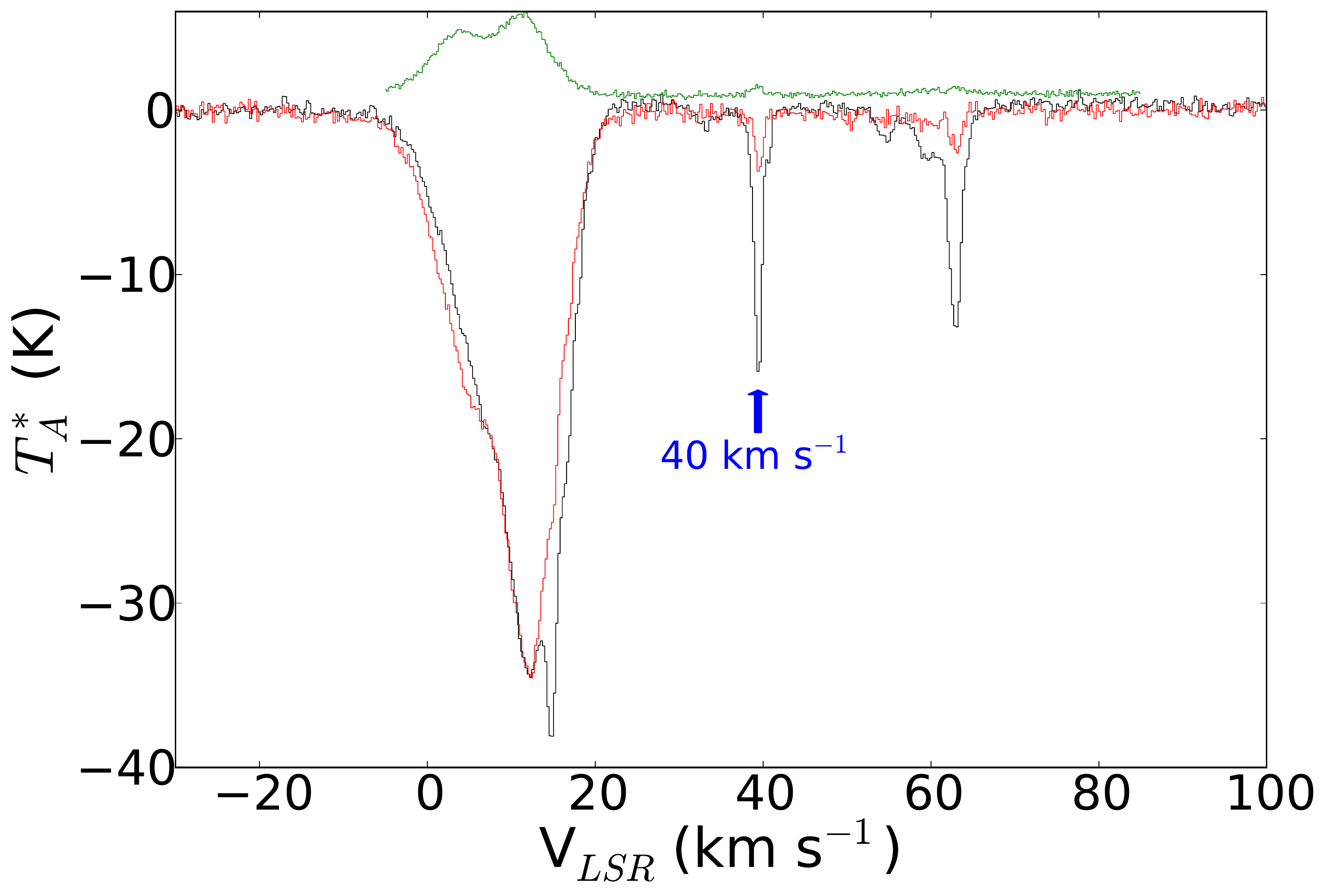}
          {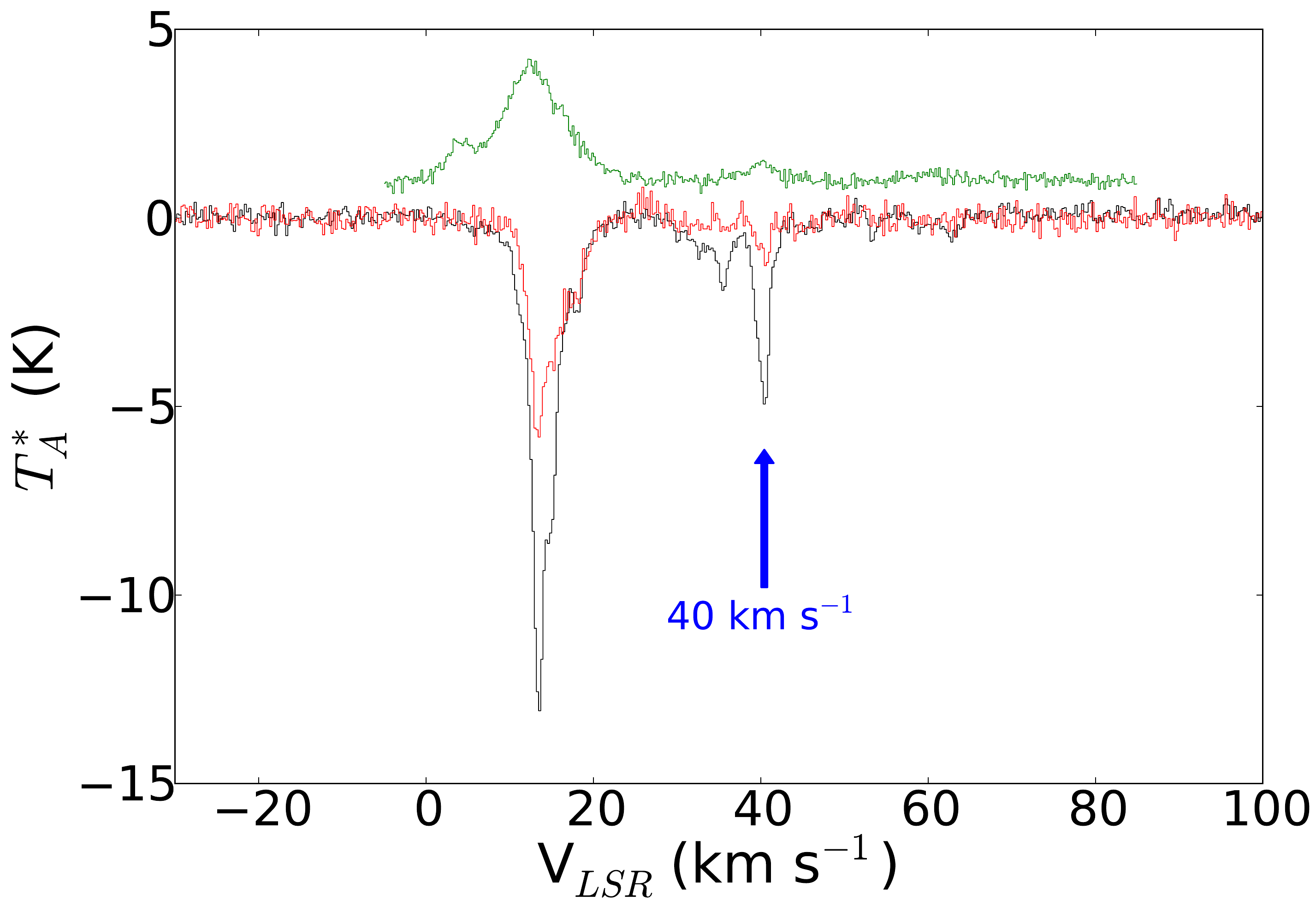}
{Spectra of the \formaldehyde \oneone (black), \twotwo (red), and \thirteenco
1-0 (green) lines toward G43.17+0.01 (left) and G43.16-0.03 (right).
The \formaldehyde spectra are shown continuum-subtracted, and the \thirteenco
spectrum is offset by +1 K for clarity.  The GBT \twotwo spectra are multiplied
by a factor of 9 so the smaller lines can be seen.  The blue arrow marks the 40 \kms
cloud \GRSMC that we discuss in this paper. 
}{fig:w49fullspec}{1}

We focus on the ``foreground'' line at $\sim40$ \kms, since it is not
associated with the extremely massive W49 region, which is dominated by gravity
and stellar feedback rather than pure turbulence.  The cloud is shown in Figure
\ref{fig:40kmscloud}.  
Additional \formaldehyde spectra of surrounding sources that are both bright at
8--1100 \um and within the \thirteenco contours of the cloud have \formaldehyde
\twotwo detections at $\sim 10$ or $\sim 60$ \kms. The detections of dense gas
at these other velocities, and corresponding nondetections of \twotwo at 40
\kms, indicate that the star-forming clumps apparent in the infrared in Figure
\ref{fig:40kmscloud} are not associated with the 40 \kms cloud.  

The \formaldehyde lines are observed in the outskirts of the cloud, not at
the peak of the \thirteenco emission.  The cloud spans $\sim0.6\degrees$, or
$\sim30$ pc at $D=2.8$ kpc \citep{Roman-Duval2009a}.  It is detected in \oneone
absorption at all 6 locations observed in \formaldehyde (Figure
\ref{fig:40kmscloud}), but \twotwo is only detected in front of the W49 HII
region because of the higher signal-to-noise at that location.  The detected
\thirteenco and \formaldehyde lines are fairly narrow, with \formaldehyde FWHM
ranging from $\Delta v \sim1.3-2.8$ \kms and \thirteenco widths from $\Delta v
\sim1.5-4.6$ \kms, where the largest line-widths are from averaging over the
largest scales in the cloud.  The \thirteenco lines are 50--100\% wider than the
\formaldehyde lines.  This greater linewidth is due to high optical depth in the
more common isotopologues, since \ceighteeno has the same linewidth as \formaldehyde
and \twelveco is $3\times$ wider \citep[][their Table 4]{Plume2004a}.

The highest \thirteenco contours are observed as a modest infrared dark cloud
in Spitzer 8 \um images, but no dust emission peaks are observed at 500 \um
\citep[Herschel;][]{Traficante2011a} or 1.1 mm
\citep[Bolocam;][]{Aguirre2011a,Ginsburg2013a} associated with the dark gas.
This is an indication that the cloud is not dominated by gravity -- no
massive dense clumps are present within this cloud.

The cloud's density is the key parameter we aim to measure, so we first determine the cloud-averaged
properties based on \thirteenco 1-0.
The cloud has mass in the range $M_{CO} = 1-3\ee{4}$ \msun in a radius $r=15$ pc as measured
from the integrated \thirteenco map using an optical depth estimate and
abundance from \citet{Roman-Duval2010a}, so its mean density is $\rho(\hh) \approx
10-30$ \percc assuming spherical symmetry (see Appendix \ref{sec:caveats}).  If we
instead assume a cubic volume, as is done in simulations, the mean density is lower by a factor $\pi/6$. 
\citet{Simon2001a} report a mass $M_{CO} = 6\ee{4} \msun$ and $r=13$ pc,
yielding a density $\rho(\hh)=100$ \percc, which is consistent with our estimates.
\citet{Roman-Duval2010a} break the cloud apart into 3 separate objects for their
analysis, GRSMC 43.04-0.11, GRSMC 43.24-00.31, and GRSMC 43.14-0.36.  All three
have the same velocity to within 1 \kms, but they show slight discontinuities
in position-velocity space.  These discontinuities are morphologically consistent
with gaps seen in turbulent simulations, validating our assessment of the cloud as a
single object, but as a maximally conservative estimate we use the density of the
northmost ``clump'' GRSMC 43.04-0.11, which overlaps our target line of sight,
as an upper limit.  It has density $\rho\approx120~\percc$, but we use $\rho<200$ \percc
as a slightly more conservative limit to allow for modest uncertainties in
optical depth, radius, and abundance.

\Figure{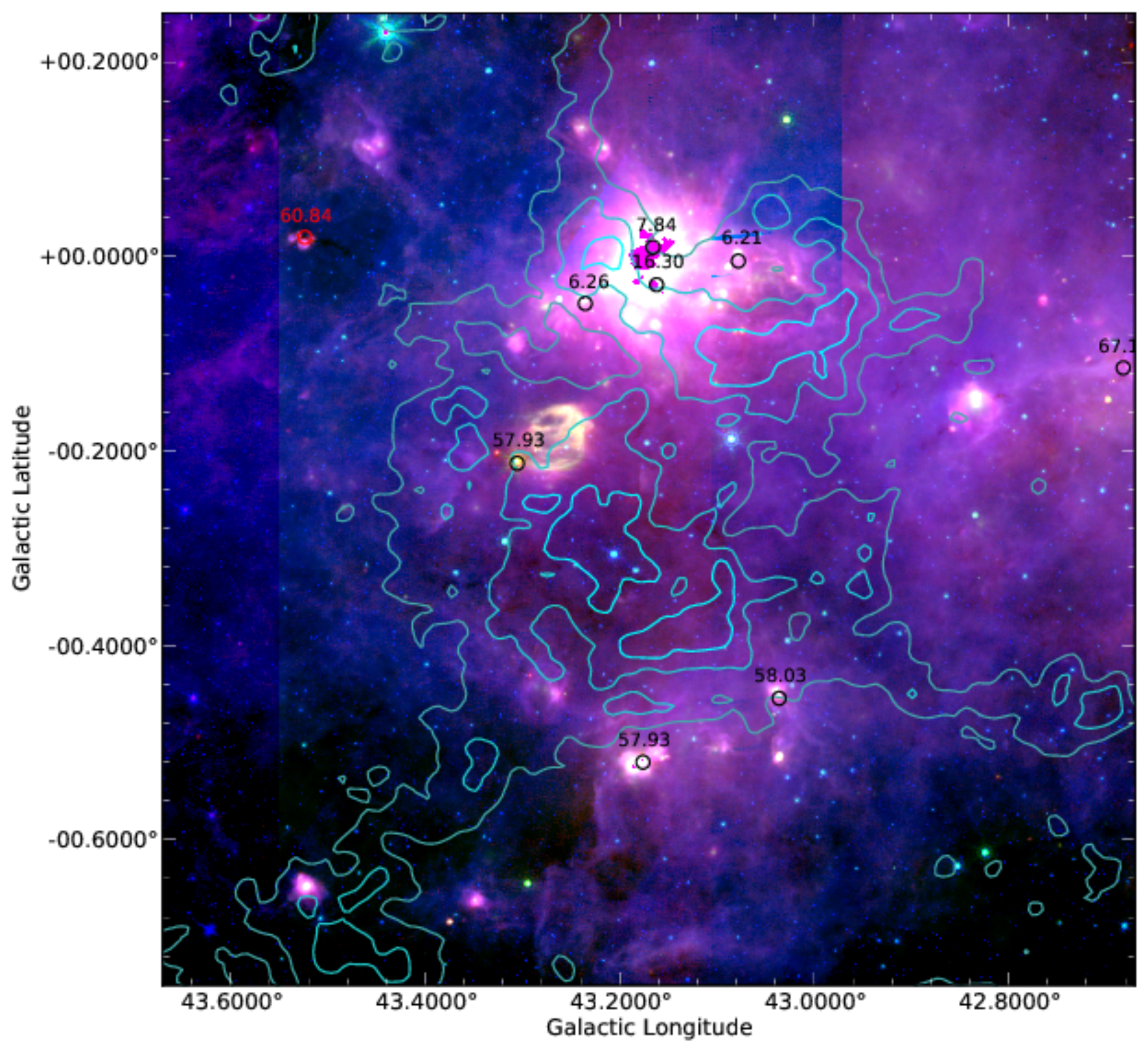}
{\textit{The GRSMC 43.30-0.33 cloud.}  The background image shows Herschel SPIRE 70 \um (red),
Spitzer MIPS 24 \um (green), and Spitzer IRAC 8 \um (blue) with
the GRS \thirteenco \citep{Jackson2006a} integrated image from $v_{LSR}=36$ \kms to $v_{LSR}=43$ \kms at contour levels of
1, 2, and 3 K \kms superposed in cyan contours.  The red and black circles
show the locations and beam sizes of the \formaldehyde observations, and their labels indicate the LSR velocity
of the deepest absorption line in the spectrum.  The W49 HII region is seen
behind some of the faintest \thirteenco emission.  The dark swath in the 8 and
24 \um emission going through the peak of the \thirteenco emission in the lower
half of the image is a low optical depth infrared dark cloud associated with
this cloud.  The two pointings examined in this paper and shown in Figures
\ref{fig:w49fullspec} and \ref{fig:h2codensg43} are labeled by their peak LSR
velocities, 7.84 and 16.30, for \north and \south respectively.  They are separated by
about 1 pc at the distance to the 40 \kms cloud.}
{fig:40kmscloud}{0.75}{0}

\section{Modeling \formaldehyde}
\label{sec:modeling}
In order to infer densities using the \formaldehyde densitometer, we use the
low-temperature collision rates given by \citet{Troscompt2009a}\footnote{The \citet{Wiesenfeld2013a}
rates provide access to higher temperatures, but for the low temperatures we are treating in this
paper, the \citet{Troscompt2009a} values are slightly more accurate (Alexandre Faure, private communication).}
with RADEX
using the large velocity gradient (LVG) approximation \citep{van-der-Tak2007a} to build a grid of
predicted line properties covering 100 densities $\rho(\hh) = 10-10^8$ \percc,
10 temperatures $T=5-50$ K, 100 column densities $N(\ortho) = 10^{11}-10^{16}$
\persc, and 10 \hh ortho-to-para ratios $OPR = 0.001-3.0$.

The \formaldehyde densitometer measurements are shown in Figure \ref{fig:h2codensg43}.
The figures show optical depth spectra, given by the equation
\begin{equation}
    \tau = -\ln\left(\frac{S_\nu + 2.73\mathrm{~K}}{\bar{C_\nu} + 2.73\mathrm{~K}}\right)
\end{equation}
where $S_\nu$ is the spectrum (with both the line and continuum included) and $\bar{C_\nu}$ is
the measured continuum, both in Kelvin.  The cosmic microwave background
temperature is added to the continuum since \formaldehyde can be seen in
absorption against it, though toward W49 it is negligible.

Since the W49 lines of sight are clearly on the outskirts of the foreground
cloud, not through its center, it is unlikely that these lines of sight
correspond to a centrally condensed density peak (e.g., a core).  The
comparable line ratios observed through two different lines of sight separated
by $\sim 1$ pc supports this claim, since if either line was centered on a
core, we would observe a much higher \twotwo optical depth.

\FigureTwo{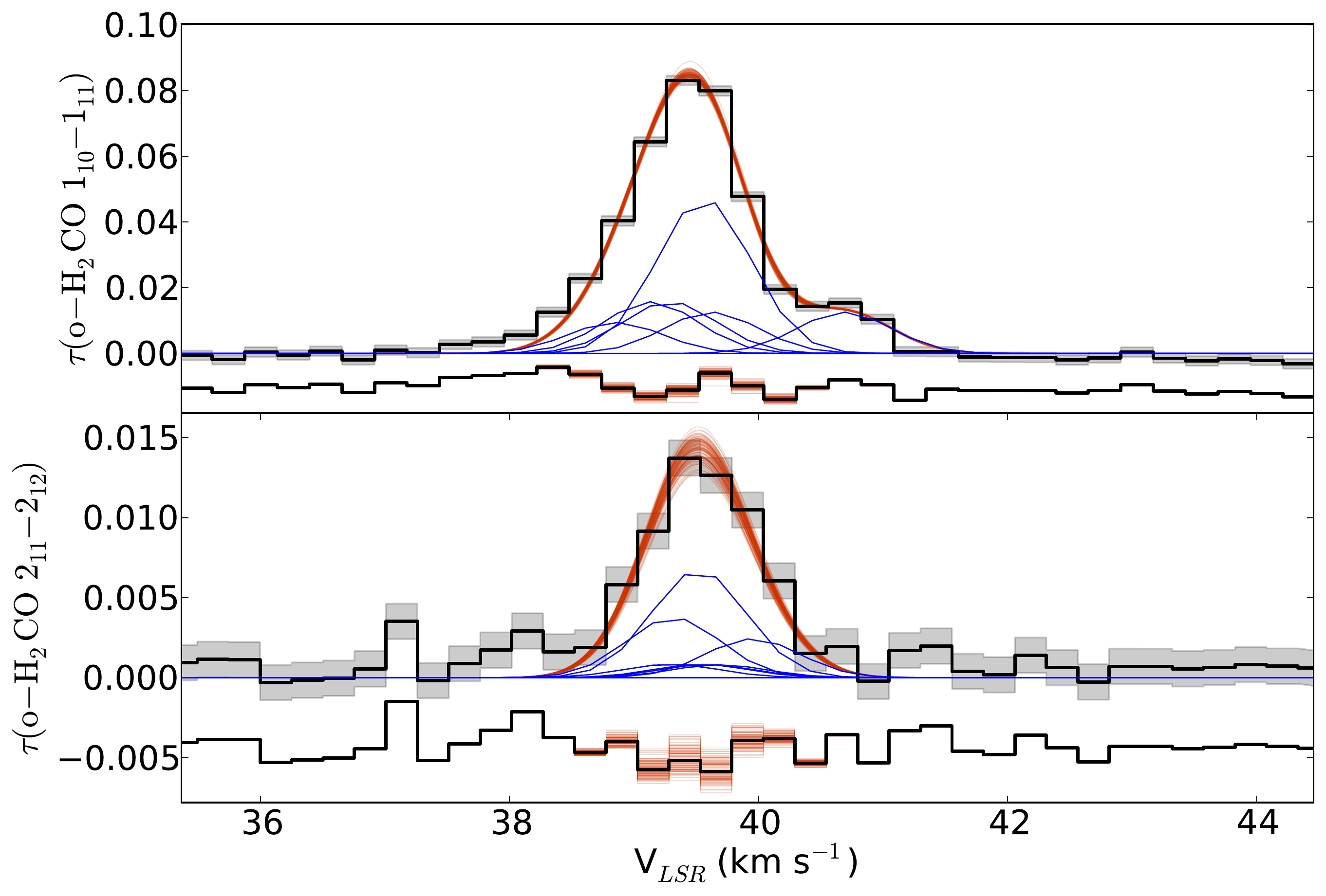}
          {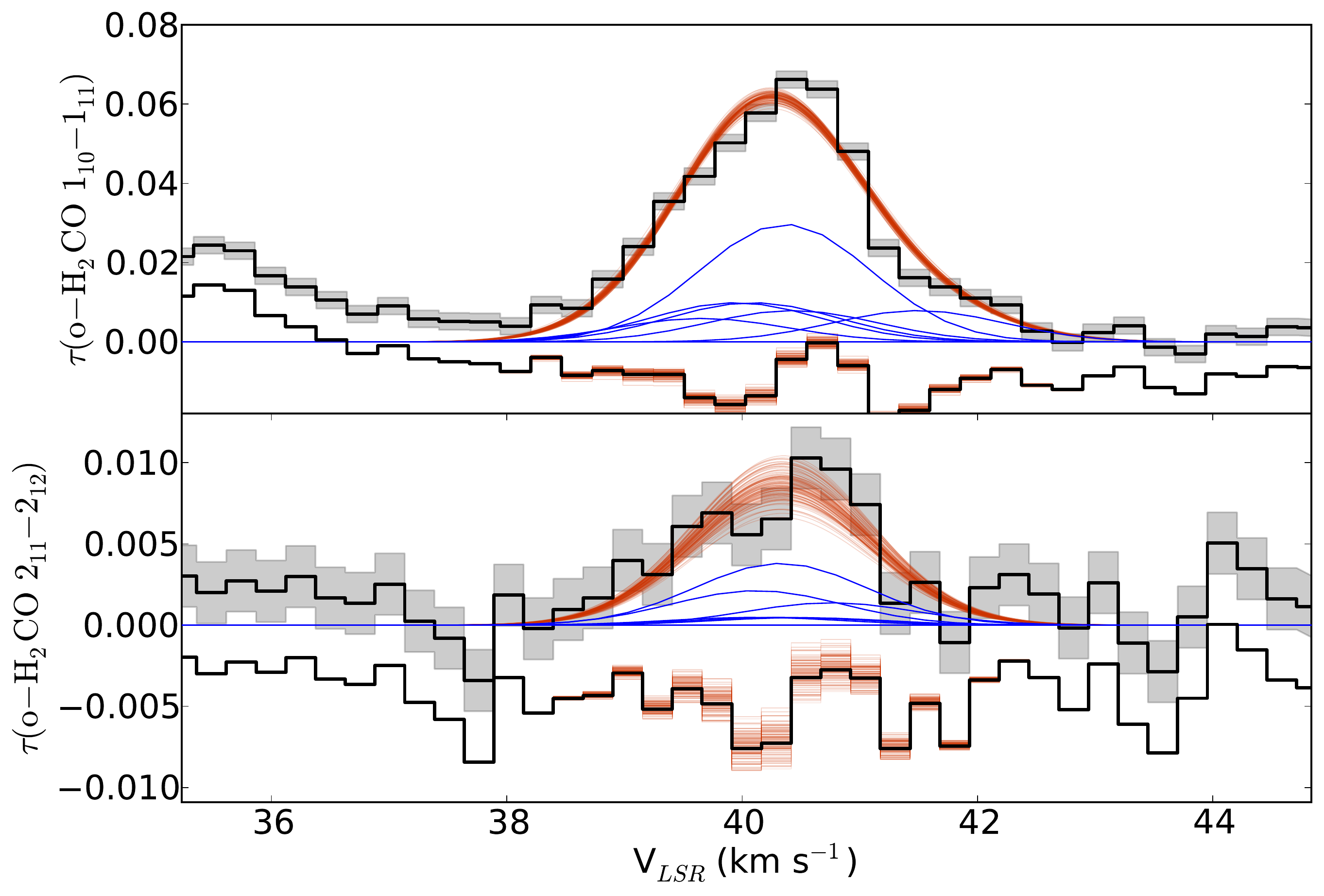}
{Optical depth spectra of the \oneone and \twotwo lines toward the two W49
lines of sight, \north (left) and \south (right).   The grey bars show the
1-$\sigma$ error bars on each data point.  The red lines show 100 realizations
from an MCMC fit of the \ortho \oneone and \twotwo lines using the LVG model
grid.  The blue lines show the hyperfine components that make up the \oneone
and \twotwo lines for the optimal fit; the \oneone line is resolved into two
components in the \north spectrum.  The residuals of the fit are shown offset
below the spectra with the residuals of the above 100 MCMC realizations
overplotted in red.  The optical depth ratio falls in a regime where
gas temperature has very little effect on the observed depth and there is no
degeneracy between low and high densities \citep{Ginsburg2011a}.  }
{fig:h2codensg43}{1}

We performed fits of the optical depth spectra to each line independently using
a  Markov-Chain Monte Carlo (MCMC) approach \citep{Ginsburg2011c,Patil2010a}. In
both lines of sight, we found that the centroids and widths agreed (see Table
\ref{tab:obs}).  

From this point on, we discuss only the \north line of sight ($V_{LSR}=7.84$
\kms in Figure \ref{fig:40kmscloud}), since it is well-fit by a single
component and has high signal-to-noise.  Since both lines of sight
sample the same CO cloud, all of the measurements below are most strongly
constrained by the \north line of sight and the \south line of sight provides
no additional information.

\Table{lcc}
{Fitted Parameters}
{  & \oneone & \twotwo}
{tab:obs}
{
\north    &                          &                         \\
\hline
Centroid  & $  39.54^{+   0.01}_{-   0.01}$  & $  39.55^{+   0.06}_{-   0.06}$ \\
Width     & $   0.37^{+   0.01}_{-   0.02}$  & $   0.45^{+   0.07}_{-   0.08}$ \\
Peak      & $  0.114^{+  0.004}_{-  0.004}$  & $  0.015^{+  0.002}_{-  0.002}$ \\
Integral  & $  0.107^{+  0.002}_{-  0.002}$  & $  0.016^{+  0.002}_{-  0.002}$ \\
Ratio     & \multicolumn{2}{c}{$   6.49^{+   0.84}_{-   0.67}$}  \\
\hline
\south    &                            &                         \\
\hline
Centroid  & $  40.35^{+   0.04}_{-   0.03}$  & $  40.36^{+   0.23}_{-   0.22}$ \\
Width     & $   0.72^{+   0.04}_{-   0.04}$  & $   0.84^{+   0.23}_{-   0.31}$ \\
Peak      & $  0.071^{+  0.003}_{-  0.003}$  & $  0.008^{+  0.002}_{-  0.002}$ \\
Integral  & $  0.130^{+  0.005}_{-  0.005}$  & $  0.018^{+  0.004}_{-  0.004}$ \\
Ratio     & \multicolumn{2}{c}{$   7.32^{+   2.31}_{-   1.43}$}  \\
}{
Centroid and width are in \kms, peak is unitless (optical
depth), and the integral is in optical depth times \kms.  The errors represent
95\% credible intervals (2-$\sigma$).}

\section{Turbulence and the \formaldehyde cm lines}
\label{sec:turbulenceh2co}
Supersonic interstellar turbulence can be characterized by its driving mode,
Mach number $\mathcal{M}$, and magnetic field strength. 
We start by assuming the gas density follows a lognormal distribution, defined
as 
\begin{equation}
    \label{eqn:lognormal}
    P_V(s) = \frac{1}{\sqrt{2 \pi \sigma_s^2}} \exp\left[-\frac{(s+\sigma_s^2/2)^2}{2 \sigma_s^2}\right]
\end{equation}
\citep{Padoan2011b,Molina2012a}
where the subscript $V$ indicates that this is a volume-weighted density
distribution function.  The parameter $s$ is the logarithmic density contrast,
$s\equiv\ln(\rho/\rho_0)$ for mean volume-averaged density $\rho_0 \equiv \meanrho_V$.
The width of the turbulent density distribution
is given by
\begin{equation}
    \label{eqn:sigmas}
    \sigma_s^2 = \ln\left(1+b^2 \mathcal{M}^2 \frac{\beta}{\beta+1}\right)
\end{equation}
where $\beta= 2 c_s^2/v_A^2 = 2 \mathcal{M}_A^2/\mathcal{M}^2$ and $b$ ranges
from $b\sim1/3$ (solenoidal, divergence-free forcing) to $b\sim1$ (compressive,
curl-free) forcing \citep{Federrath2008a,Federrath2010a}.  $c_s$ is the
isothermal sound speed ($s$ here is short for `sound'), $v_A$ is the Alfvén
speed, and $\mathcal{M}_A$ is the Alfvénic Mach number.

The observed \formaldehyde line ratio roughly depends on the
\emph{mass-weighted} probability distribution function (as opposed to the
volume-weighted distribution function, which is typically reported in
simulations).  For each \formaldehyde molecule, the likelihood of absorbing a
background photon is set by the level population in the lower energy state,
which is controlled by the \hh density as long as the line is optically thin
(which is the case we treat here).

For a given `cell' at density $\rho$, the optical depth is given by the number
(or mass) of particles in that cell $M(\rho) = V \cdot \rho$  (assuming a fixed
cell volume $V$) times the optical depth $\Upsilon_{\nu,p}$, where the subscript $p$ indicates
that this is an optical depth per particle.
The total optical depth is the optical depth per cell integrated over the
probability distribution function, $\tau_{tot} = \int M(\rho) \Upsilon_{\nu,p}
P_V(\rho) d\rho$, which is equivalent to $\tau_{tot} = \int \Upsilon_{\nu,p} P_M(\rho) d\rho$
using the definition of mass-weighted density $P_M(\rho) \equiv (\rho/\rho_0) P_V(\rho)$.

Following this derivation,
we use the RADEX models of the \formaldehyde lines, which are computed assuming a
fixed local density, as a starting point to model the observations of
\formaldehyde in turbulence.   Starting with a fixed \emph{volume-averaged}
density $\rho_0$, we compute the observed \formaldehyde optical depth $\tau_\nu$ in both
the \oneone and \twotwo
line by averaging over the mass-weighted density distribution and redefining the equations with
a logarithmic differential. 
\begin{eqnarray}
    \label{eqn:tauintegral}
    \tau_{\nu}(\rho_0) &=& \int_{-\infty}^\infty \Upsilon_{\nu,p}(\rho) P_M(\ln \rho/\rho_0) d \ln (\rho/\rho_0)\\
                       &=& \int_{-\infty}^\infty \Upsilon_{\nu,p}(\rho_0 e^s) P_M(s) d s
\end{eqnarray} 
$\Upsilon_{\nu,p}(\rho)$ is the optical depth \emph{per particle} at a given density, where $N_p$ is the column
density (\perkmspc) from the LVG model.
We assume a fixed abundance of \ortho relative to \hh
(i.e., the \formaldehyde perfectly traces the \hh).\footnote{While there is
building evidence that there is \hh not traced by CO
\citep{Glover2010a,Shetty2011b,Shetty2011a},  \formaldehyde abundances have
typically been observed to be
consistent with CO abundances, so the mass traced by the CO is the same
we observe in \formaldehyde.  \formaldehyde deficiency is also most likely to
occur on the optically thin surfaces of clouds where the total gas density is
expected to be lower, so our measurements should be largely unaffected by
abundance variation within the cloud.}

Figure \ref{fig:lvgsmooth}
shows the result of this integral for an abundance of \ortho relative to \hh, 
$X(\ortho)=10^{-8.5}$, where the X-axis shows the volume-averaged number density $\rho_0 \equiv \rho(\hh)$ and the Y-axis
shows the observable optical depth ratio of the two \formaldehyde centimeter
lines.
The LVG model, which assumes a single density
(or a Dirac $\delta$ function as the density distribution), is shown along with
the PDF-weighted-average versions of the model that take into account realistic turbulent
gas distributions.  

The \formaldehyde \twotwo line requires a higher density to be ``refrigerated''
into absorption.  As a result, any spread of the density distribution means
that a higher fraction of the mass is capable of exciting the \twotwo line.
Wider distributions increase the \twotwo line more than the \oneone line and
decrease the (\oneone)/(\twotwo) ratio.

\FigureFourPDF
    {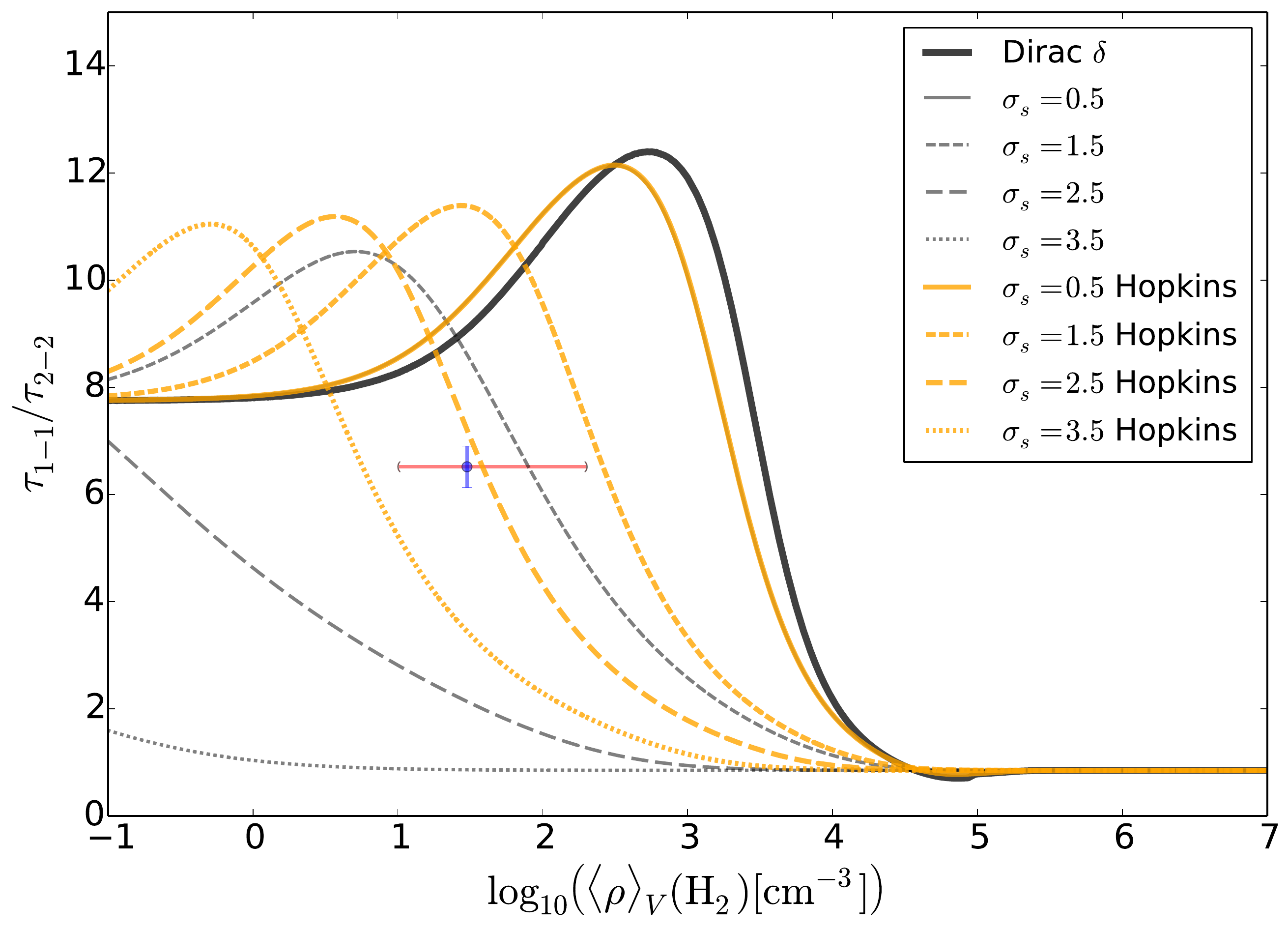}
    {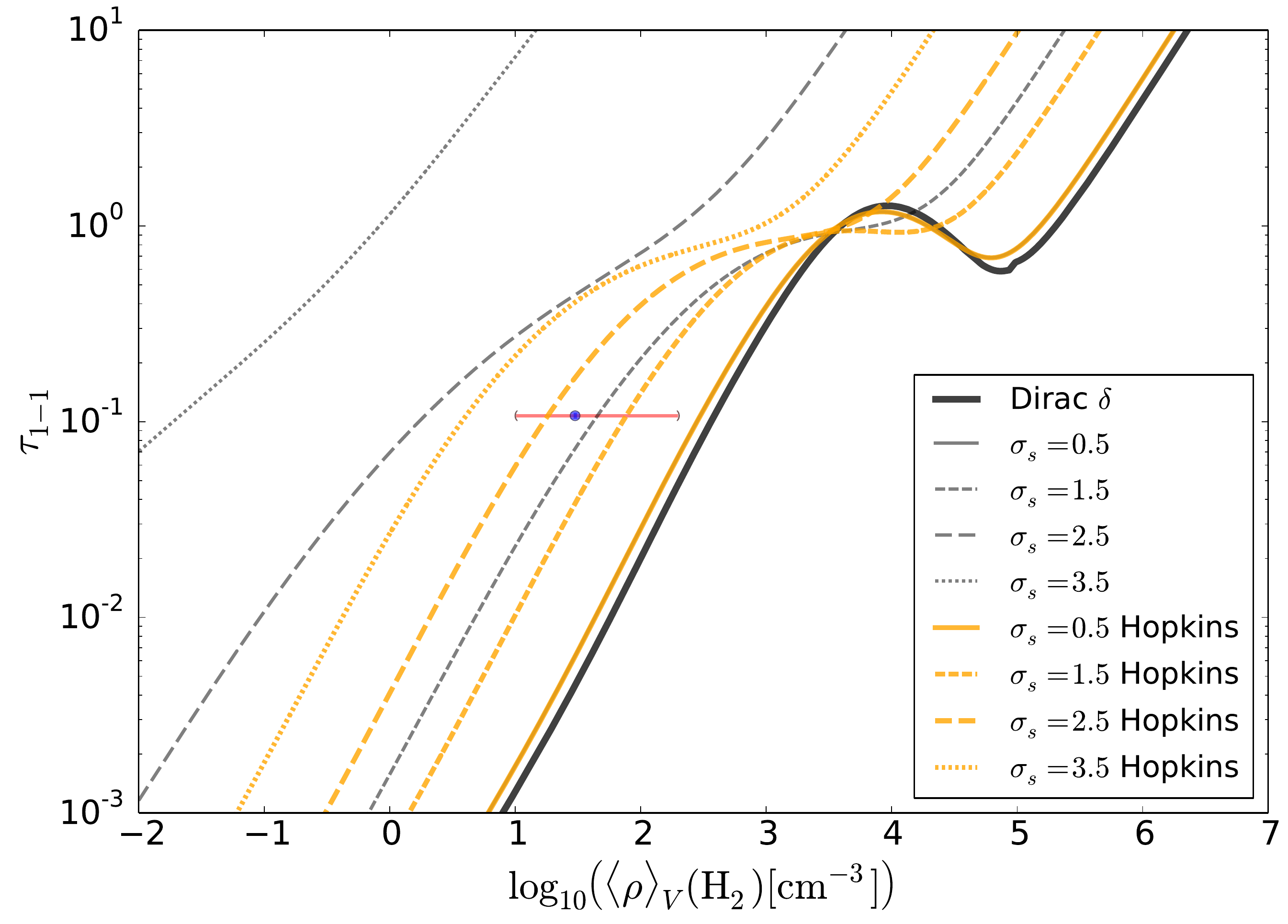}
    {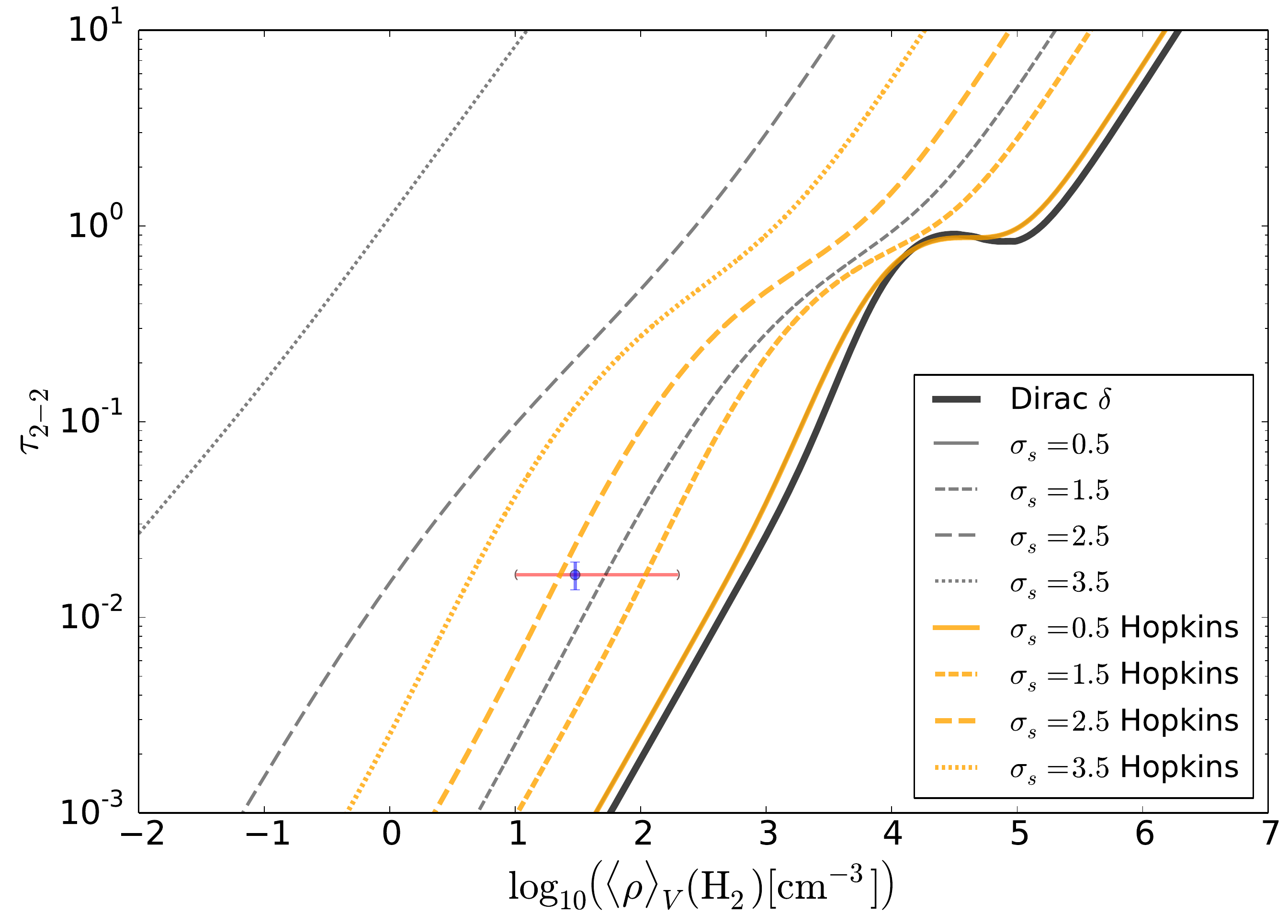}
    {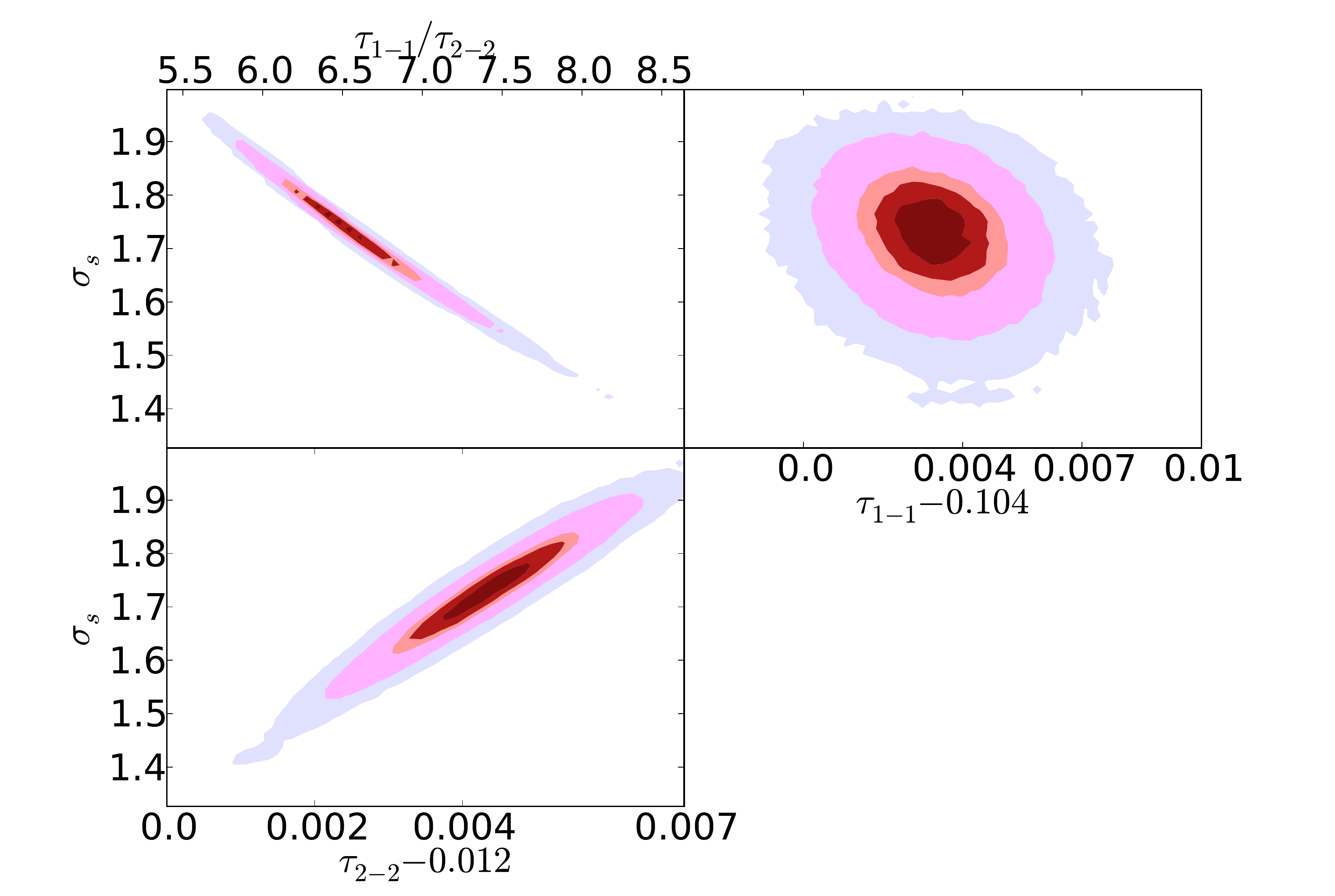}
{The predicted \formaldehyde \oneone/\twotwo ratio and optical depths as a
function of the \thirteenco-derived volume-weighted mean density for a fixed
abundance relative to \hh $X(\ortho) = 10^{-8.5}$  with \hh ortho/para ratio
1.0.  The different lines show the effect of averaging over different
mass distributions as identified in the legend. 
The thick solid line shows the predicted values
with no averaging (i.e., a $\delta$-function density distribution); the other solid line
shows $\sigma_s=0.5$ for both distributions (they overlap).
The blue point shows the \north measurement.  The horizontal red error bars
show the limits on the mean volume density, $\meanrho_V$, and the vertical blue error
bars show the 95\% credible interval for the \formaldehyde line measurements.
The bottom-right figure shows the allowed $\sigma_s$ parameter space for the
lognormal distribution given the \oneone and \twotwo measurements and their
ratio; the values are reported in Table \ref{tab:datafits}.  The contours
indicate the 25\% (dark red), 50\% (red), 68\% (light red), 95\% (pink), and 99.7\%
(blue-grey) credible regions.
}
{fig:lvgsmooth}{3.5in}

\subsection{The $\rho$-PDF in \GRSMC}
\label{sec:grsmcturb}
We use the density measurements in \GRSMC to infer properties of that
cloud's density distribution.  The observed line ratio for the \north sightline in
\GRSMC is shown in Figure \ref{fig:lvgsmooth} as a blue point.  The position
of this point on the x-axis is set by the \thirteenco-derived volume-averaged density,
while its y-axis position in the three subplots reflects the \formaldehyde measurements
reported in Table \ref{tab:obs}.

\Table{ccccc}
{Fitted Distribution Parameters}
{Parameter & Lognormal & Hopkins}
{tab:datafits}
{
$X(\formaldehyde)=10^{-8.5}$ \\
\hline
$\sigma_s$&1.7$^{0.2}_{0.2}$& - \\
$\sigma_s | \mathcal{M}$&1.7$^{0.2}_{0.1}$& - \\
$b|\mathcal{M}$&$>0.56$& -  \\
\hline
$X(\formaldehyde)=10^{-9.0}$ \\
\hline
$\sigma_s$&1.5$^{0.1}_{0.1}$&2.7$^{0.5}_{0.5}$ \\
$T$&-&0.31$^{0.08}_{0.10}$ \\
$\sigma_s | \mathcal{M}$&1.5$^{0.1}_{0.1}$&2.5$^{0.5}_{0.5}$ \\
$T|\mathcal{M}$&-&0.29$^{0.08}_{0.08}$ \\
$b|\mathcal{M}$&$>0.41$&$>0.71$ \\
}
{\\ The error bars represent 95\% credible intervals.  For the $b$ parameter, only
the lower limit is shown.  The $|\mathcal{M}$ notation indicates that the
parameter measurement includes the constraints imposed by the Mach number
measurements, for which we have adopted $\mathcal{M}_{3D} = 5.1 \pm 1.5$, where
$\sigma_{\mathcal{M}}=1.5$ is the standard deviation of the normal distribution
we used to represent the Mach number.  The $-$'s indicate disallowed parameter
space (top) or parameters that are not part of the distribution (bottom).
}

Assuming the thermal dominates the magnetic pressure ($\beta>>1$), we can fit
$\sigma_s$ from the model distributions in Figure \ref{fig:lvgsmooth}.  Using two
different forms for the density distribution, and using only the $\tau$
measurements as a constraint, we derive the value of $\sigma_s$ in Table
\ref{tab:datafits} and seen in the bottom-right panel of Figure \ref{fig:lvgsmooth}.

Direct measurements of the Mach number from line-of-sight velocity dispersion
measurements allow for further constraints on the distribution shape.  Assuming
a temperature $T=10$ K, consistent with both the \formaldehyde and CO
observations \citep{Plume2004a}, the sound speed in molecular gas is $c_s=0.19$
\kms.  The gas is unlikely to be much colder, so this sound speed provides an
upper limit on the Mach number.  The observed line FWHM in G43.17 is 0.95 \kms
for \formaldehyde and 1.7 \kms for \thirteenco 1-0,
so the 3-D Mach number of the turbulence is \citep{Schneider2013a}
\begin{equation}
    \label{eqn:mach}
    \mathcal{M}_{3D} \equiv 3^{1/2} \mathcal{M}_{1D} \approx \frac{3^{1/3}}{(8\ln 2)^{1/2}} FWHM / c_s 
\end{equation}
or $\mathcal{M}_{3D} = 3.8 - 6.6 $ , ranging from the \formaldehyde to the \thirteenco width along
the \north line of sight.  However, we note that the velocity dispersion for the whole cloud is
larger.

\FigureTwo{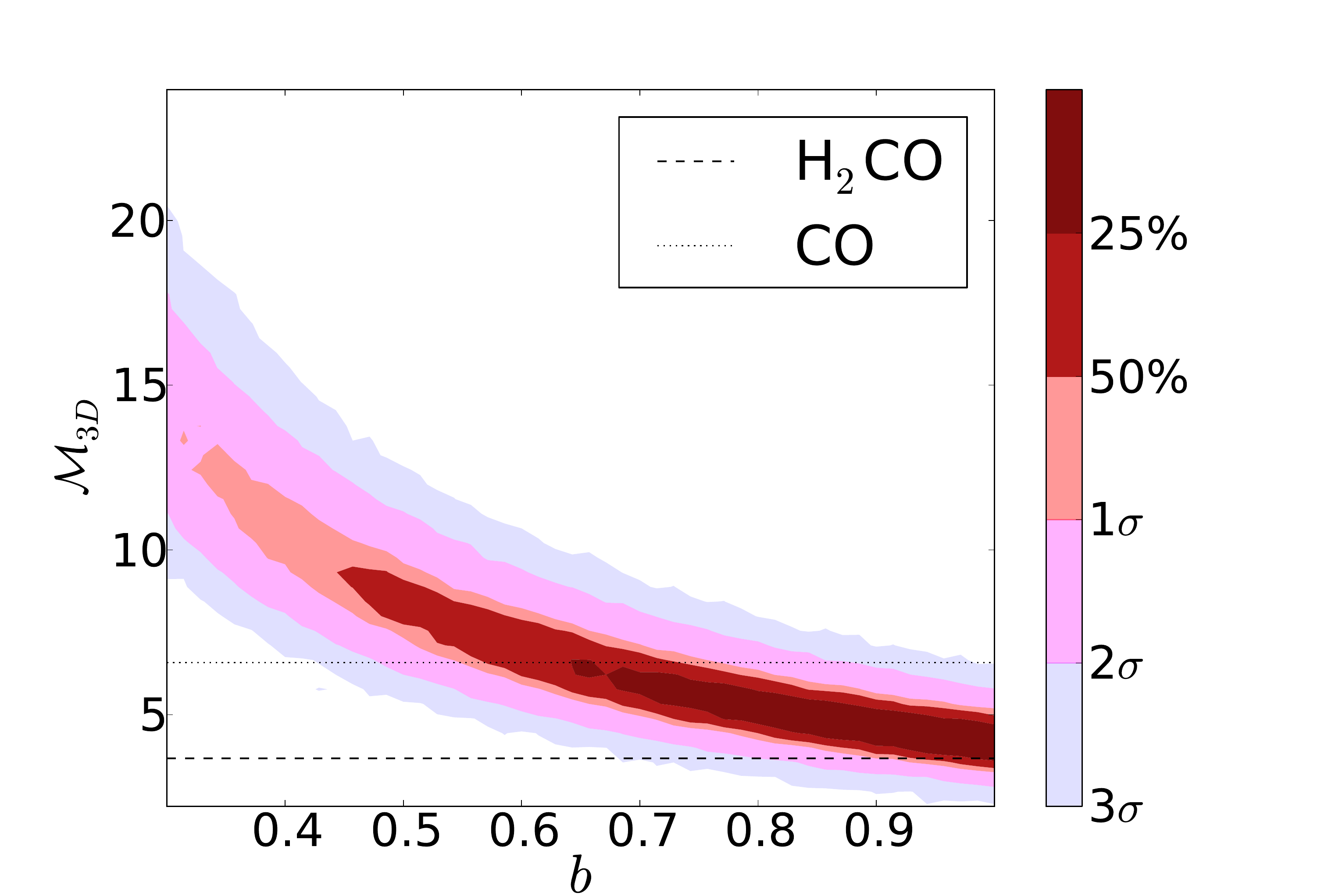}
          {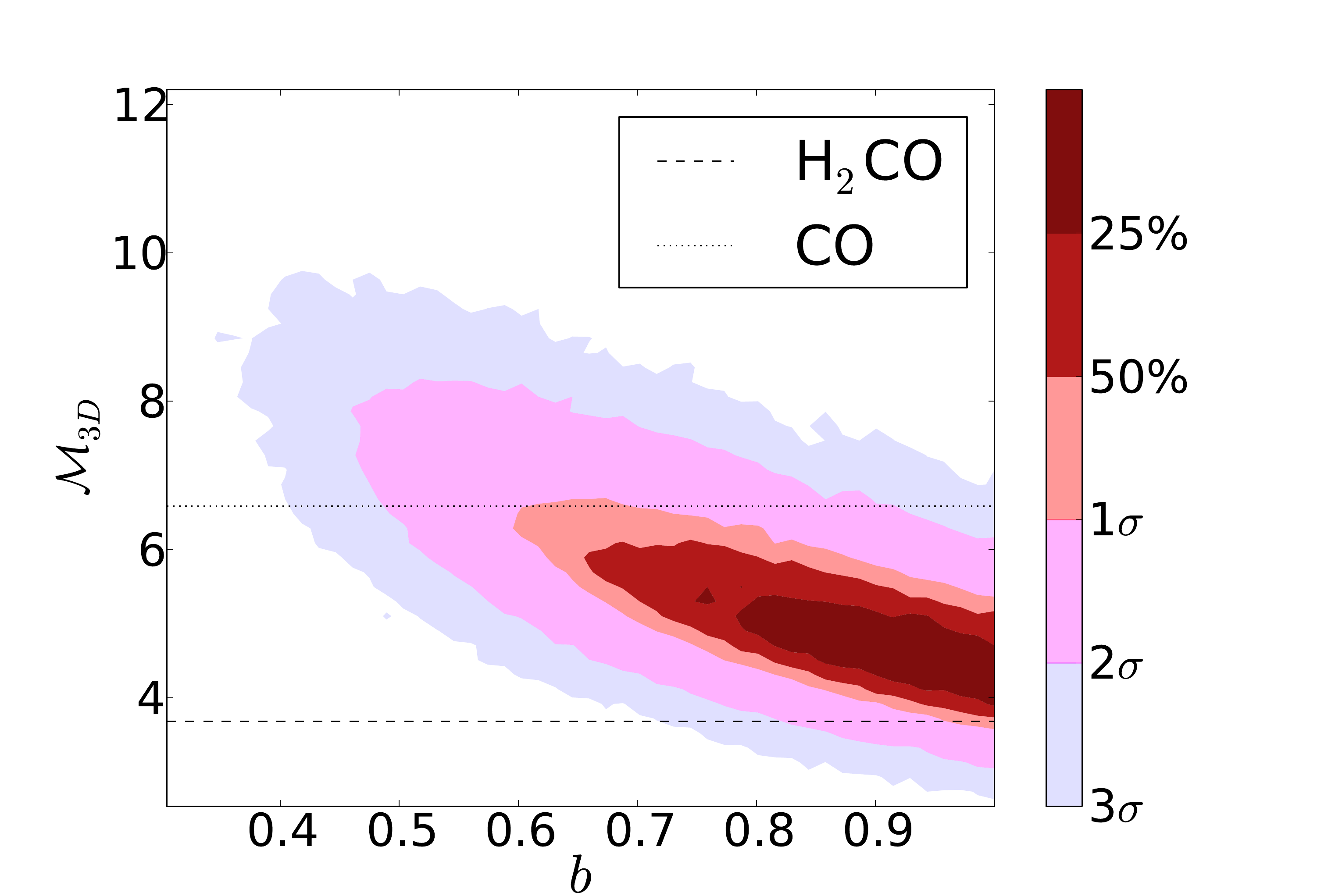}
{Contours of the MCMC fit to the \formaldehyde optical depths with the cloud
mean density restricted to $10~\percc < \meanrho_V < 200~\percc$.  The contour
levels indicate the regions in which 25, 50, 68, 95, and 99.7\% of the MCMC
samples are included. The left plot shows the parameter space allowed with no
constraints on the Mach number, indicating the mild degeneracy between Mach and
$b$.  The right plot shows the parameters derived using the constraints on the
Mach number based on the \north line of sight, $\mathcal{M}_{3D}\approx5.1\pm1.5$, which is
halfway between the value inferred from the \formaldehyde and \thirteenco line widths.  
The horizontal lines in both plots represent the Mach numbers inferred from the
\formaldehyde and CO line widths via Equation \ref{eqn:mach}.  Both plots are marginalized
over the other free parameters ($\sigma_s$, $\rho_V$, and the observed optical depth).
}{fig:lognormalmcmc}{1}

Using the observed range of Mach numbers along the \north line of sight, we can
constrain $b$ with Equation \ref{eqn:sigmas}.  Figure \ref{fig:lognormalmcmc} shows the Mach number-$b$
parameter space allowed by the observed volume density and \formaldehyde lines
both with and without the Mach number constraint imposed.  If we assume the Mach number
is approximately halfway between the \formaldehyde and CO based measurements, with
a dispersion that includes both, we can constrain $b>0.56$ (see Table \ref{tab:datafits}).

\subsubsection{The Hopkins distribution}
As one possible alternative, we use the
\citet{Hopkins2013a} density distribution,
\begin{equation}
    \label{eqn:hopkins}
    P_V (\ln \rho) d \ln \rho = I_1 (2 \sqrt{\lambda u}) e^{-(\lambda+u)} \lambda du
\end{equation}
where $u \equiv \lambda/(1+T) - \ln(\rho/\rho_0)/T$ and $\lambda\equiv \sigma^2_{\ln \rho/\rho_0} / (2 T^2)$ \citep[Equation 5
in ][modified such that $\rho_0$ is not assumed to be unity]{Hopkins2013a}.
The distribution is governed by a width
$\sigma_s \equiv \sigma_{\ln \rho/\rho_0}$ and an ``intermittency'' parameter $T$ that indicates the deviation of
the distribution from lognormal.  The intermittency parameter is described in
\citet{Hopkins2013a} as a unitless parameter which increases with Mach number
and is correlated with the strength of the deviations from the mean turbulent
properties as a function of time.  Its physical meaning beyond these simple
correlations is as yet poorly understood.

We use $T$ values given the $T-\sigma_s$ and $T-\mathcal{M}_C$
relations fitted to measurements from a series of simulations
\citep{Kowal2007a,Kritsuk2007a,Schmidt2009a,Federrath2010a,Konstandin2012a,Molina2012a,Federrath2013b}, 
where $\mathcal{M}_c$ is the compressive Mach number, e.g. $\mathcal{M}_c = b \mathcal{M}$.
The values are given by
\begin{equation} 
    \label{eqn:Tsigma}
    T(\sigma_s) = 0.25 \ln (1+0.25 \sigma_s^4 \left(1+T(\sigma_s)\right)^{-6})
\end{equation}
Equation \ref{eqn:Tsigma}
is a transcendental equation, so we use root-finding to determine $T$.  

Assuming the same abundance as above, $X(\formaldehyde)=10^{-8.5}$, the Hopkins
distribution is incompatible with our observations for the $T-\sigma$ relations considered
in \citet{Hopkins2013a} and the other $T$ values and relations explored in Figure \ref{fig:distributions}b.
Figure \ref{fig:distributions}a shows how the Hopkins
and lognormal distributions differ; the Hopkins distribution is more sharply
peaked and includes less gas above its peak density.  The incompatibility with
our observations arises because the Hopkins distribution produces lower
mass-weighted densities than the lognormal. 

However, the
Hopkins distribution is compatible with our observations if a lower abundance is assumed.
Using the Hopkins distribution with $X(\formaldehyde)=10^{-9}$, we find
$\sigma_s\sim2.5$ (see Table \ref{tab:datafits}).  This value is compatible
with the observed Mach numbers.  Using the relation 
\begin{equation}
    \label{eqn:McMT}
    b \mathcal{M} = \mathcal{M}_c  \approx 20 T
\end{equation}
from \citet{Hopkins2013a} Figure 3, we can derive a lower limit $b>0.7$.
However, there is additional intrinsic uncertainty in the coefficient in
Equation \ref{eqn:McMT} that comes from fitting the relation to simulated data,
and we have not accounted for this uncertainty.  

The Hopkins distribution is compatible with our observations, but requires
relatively extreme values of the standard deviation and $b$ parameters.  
We explored a few alternate realizations of the Hopkins distribution's
$T-\sigma$ relation, with results shown in Figure \ref{fig:distributions}.
Independent of the form of the Hopkins distribution chosen, it is more
restrictive than the lognormal distribution.

\subsection{Discussion}
The restrictions on $\sigma_s$ and $b$ using either the lognormal or Hopkins density
distribution are indications that compressive forcing must be a significant, if
not dominant, mode in this molecular cloud.  However, there are no obvious
signs of cloud-cloud collision or interaction with a supernova that might
directly indicate what is driving the turbulence.

Most of the systematic uncertainties tend to require a \emph{greater} $b$
value, while we have already inferred a lower limit
that is moderately higher than others have observed \citep{Brunt2010c,Kainulainen2013a}.
Temperatures in GMCs are typically 10-20 K, and we assumed 10 K: warmer
temperatures increase the sound speed and therefore decrease the Mach number. If
the cloud is warmer, the $b$ values again must be higher to account for the measured
$\sigma_s$.  Magnetic fields similarly have the inverse effect of $b$ on
$\sigma_s$, with decreasing $\beta$ requiring higher $b$ for the same
$\sigma_s$.

The only systematic that operates in the opposite direction is the abundance of
\ortho.  Lower abundance shifts all curves in Figure \ref{fig:lvgsmooth} up and
to the right, which decreases $\sigma_s$ and therefore allows for a lower $b$
for fixed Mach number.  However, abundances lower  than $X=10^{-9}$ are rarely
observed except in Galactic cirrus clouds \citep{Turner1989a} and highly
shocked regions like the cirumnuclear disk around Sgr A* \citep{Pauls1996a}, so
the measurements in Table \ref{tab:datafits} should bracket the allowed values.
While we only explored two possible abundances in detail, note that the
$\sigma_s$ values derived from the lognormal distribution vary little over
half-dex changes in abundance (Table \ref{tab:datafits}), indicating that this
measurement at least is robust to abundance assumptions.

We explore these caveats and others in more detail in the Appendix.

\FigureTwo{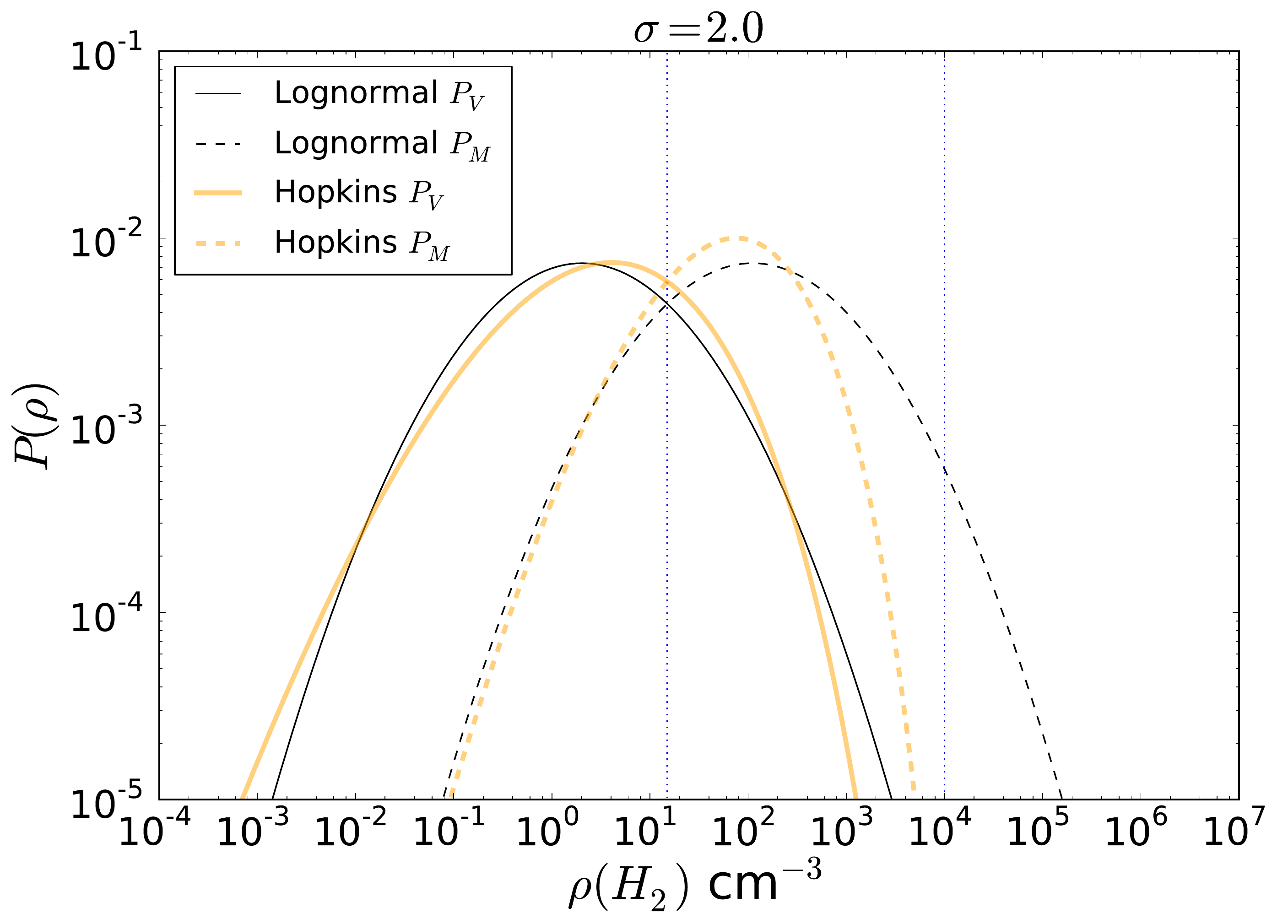}
{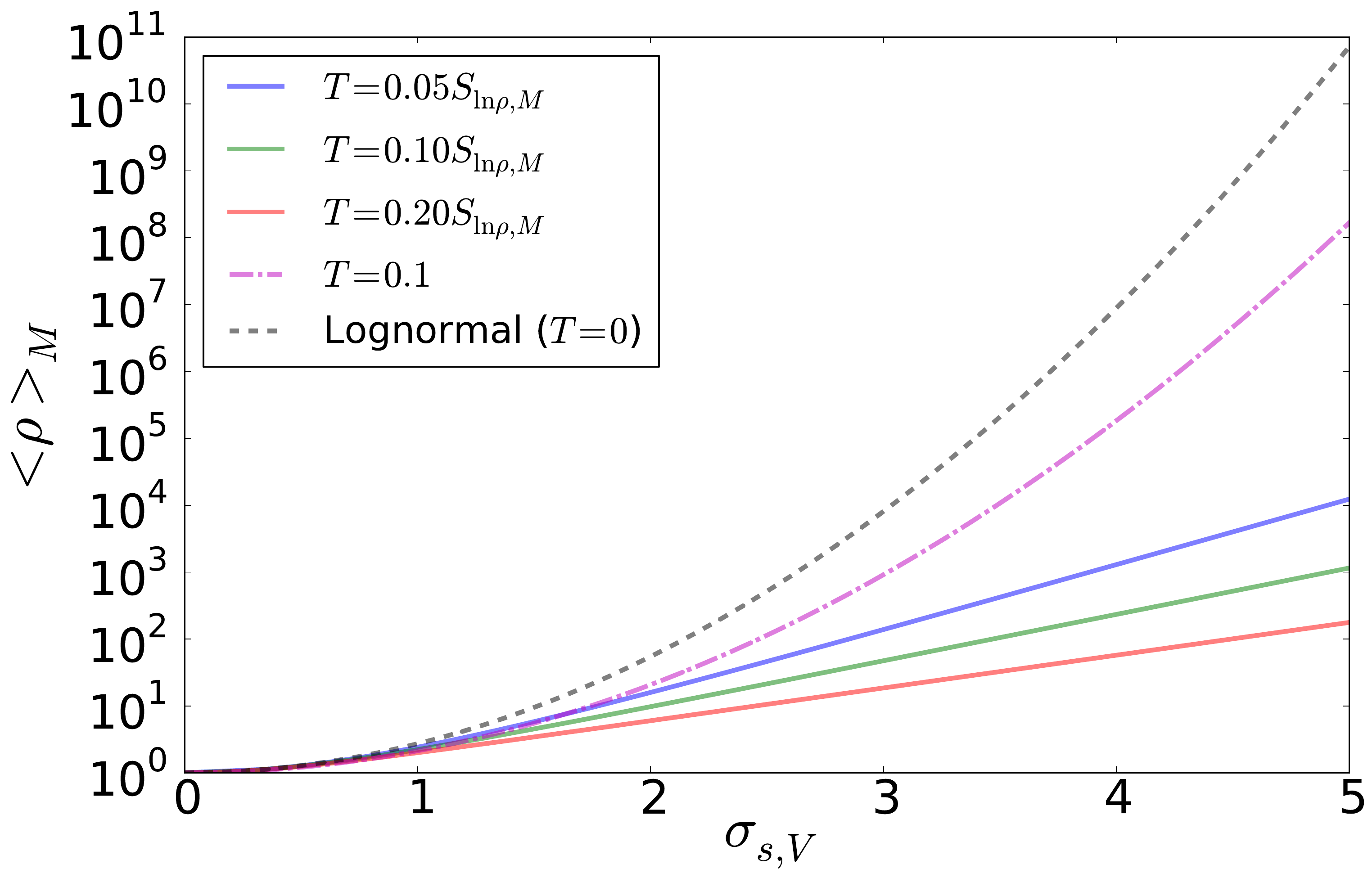}
{ 
\textit{Left:} Example volume- and mass-weighted density distributions with
$\sigma_s=2.0$.  The vertical dashed lines show $\rho = 15$ \percc and $\rho=10^4$ \percc,
approximately corresponding to the volume-weighted mean density $\meanrho_V$ of \GRSMC from \thirteenco
and the \formaldehyde-derived density, respectively.  Note that the peaks of the distributions do not correspond to their
means, since the mean of the lognormal distribution depends on its variance.
\textit{Right:} The relationship between the mass-weighted mean density
$\meanrho_M$ and the width of the volume-weighted density PDF $P_V(\rho)$ for the
lognormal distribution and different realizations of the \citet{Hopkins2013a}
distribution with $\rho_0=1$.  We show different forms of the $T-\sigma$
relation using $T=c \cdot S_{\ln \rho,M}$, which is an approximation of
Equation \ref{eqn:Tsigma}, and one example of $T=$ constant.  It is clear that, for a
given distribution width, the Hopkins distribution always puts less mass at the
highest densities than the lognormal
distribution.}
{fig:distributions}{1}

\section{Conclusions}
We demonstrate the use of a novel method of inferring parameters of the density
probability distribution function in a molecular cloud using \formaldehyde densitometry
in conjunction with \thirteenco-based estimates of total cloud mass.  We have
measured the standard deviation of the lognormal turbulence density
distribution $1.5 < \sigma_s < 1.9$ and placed a lower limit on the
compressiveness parameter $b>0.4$.  Both measurements are robust to the assumed
cloud density, \formaldehyde abundance, and other assumptions.

Our data show evidence for compressively driven turbulence in a
non-star-forming giant molecular cloud.  Since this cloud represents a typical
molecular cloud, it is likely that compressive driving is a common feature of
all molecular clouds.

This new method opens the possibility of investigating the drivers of
turbulence more directly, e.g.\ by measuring the shape of the density PDF both
within spiral arms and in the inter-arm regions.  The main requirements for
applying this technique are a moderately accurate measurement of the mean cloud
density, which can easily be provided by \thirteenco surveys such as the GRS,
and a high signal-to-noise measurement of the 2 cm and 6 cm \formaldehyde
lines.  

A precise measurement of the Mach number of the cloud will allow
\emph{measurements} of $b$ rather than the limits we have presented here.
Investigations of the predicted observed velocity dispersion and line strengths
in both \formaldehyde and \thirteenco in simulations of turbulent clouds should
provide the details needed to take this next step.

Finally, the general approach of accounting for a density distribution by
averaging over the contribution to the line profile at each density should be
generally applicable to any molecular line observations as long as the lines
are optically thin.

\section{Acknowledgements}
We thank Erik Rosolowsky and Ben Zeiger for useful discussions and comments
that greatly improved the clarity of the paper.  We thank the anonymous referee
for a timely and constructive report that helped improve the paper.
This publication makes use of molecular line data from the Boston
University-FCRAO Galactic Ring Survey (GRS). The GRS is a joint project of
Boston University and Five College Radio Astronomy Observatory, funded by the
National Science Foundation under grants AST-9800334, AST-0098562, \&
AST-0100793.
This research made use of Astropy, a community-developed core Python package
for Astronomy \citep[\url{http://astropy.org}; ][]{Astropy2013a}.  It also made use of \texttt{pyspeckit}, an
open source python-based package for spectroscopic data analysis
\citep[\url{http://pyspeckit.readthedocs.org}; ][]{Ginsburg2011c}.
The dendrogramming was performed using the python dendrogram package described at
\url{http://dendrograms.org}.
JD acknowledges the support of the NSF through the grant AST-0707713.
CF is supported from the Australian Research Council with a Discovery Projects Fellowship (Grant DP110102191).

{\it Facilities:} \facility{GBT}, \facility{Arecibo},
\facility{FCRAO}, \facility{CSO}

\appendix
\section{Assumptions, caveats, and uncertainties}
\label{sec:caveats}
We explore the various caveats and assumptions that have been treated above in
more detail here.

The precise density measurements presented here are based on large velocity
gradient approximations \citep{Sobolev1957a} for the escape probability of line
radiation from the cloud.  This method is widely used but remains an
approximation.  In the case of \formaldehyde, it has been tested with a variety
of codes \citep{van-der-Tak2007a,Henkel1983a} but is subject to uncertainties in
the velocity gradient and system geometry.  However, in the case of the
observations in this paper, the lines were observed in the optically thin regime, and the
LVG approximation should not affect our results.

The collision rates of \formaldehyde with p-\hh, o-\hh, and He are estimated
based on computer simulations of the particles.  \citet{Troscompt2009a}
improved upon the measurements of \citet{Green1991a}, bringing the typical
collision rate uncertainty down from $\sim50\%$ in the He-based approximation
to $\sim10\%$ using full models of ortho and para \hh.  \citet{Wiesenfeld2013a}
noted that the differences they observed from the \citet{Troscompt2009a} results
were $<10\%$, indicating that the methods they use are at least convergent \&
self-consistent to within $\sim30\%$.
\citet{Zeiger2010a}
reported the results of using modified collision rates assuming a 50\% error
and noted that the resulting errors in the \hh density were, in the worst case,
$<0.3$ dex.  With the improved \citet{Troscompt2009a} collision rates, the
model uncertainties are no longer dominated by collision rate uncertainties.

Abundance remains a serious concern, as most studies of \ortho abundance
do not observe multiple transitions and therefore do not constrain the relative
level populations.  There
are also general difficulties in measuring absolute abundance of molecules, as
the absolute column of \hh is rarely known with high accuracy.  Most abundance
measurements are above $X_{\ortho}>10^{-9}$
\citep{Dickens1999a,Liszt2006a}, 
except near Sgr A* \citep{Pauls1996a} and in Galactic cirrus clouds
\citep{Turner1989a,Turner1993a} where it is generally observed to have
$10^{-10}<X_{\ortho}<10^{-9}$.  These measurements dictated the abundance
boundaries we used in our analysis.

The ortho-to-para ratio of \hh is a significant uncertainty in the models,
since para-\hh is more effective at ``refrigerating'' the \formaldehyde
molecules.  Values of the ortho-to-para ratio $>1$ favor lower densities by
$\sim0.3$ dex, but we have used these lower densities in our
analysis, and therefore our results are conservative.  However, if the
ortho-to-para ratio is in reality close to zero, the density PDF must be wider
and $b$ correspondingly higher.

The ``covering factor'' of foreground clouds in front of background
illumination sources is, in general, a major concern when performing absorption
measurements.  For the clouds presented in this work, the absorbing region
is much larger than the background, as evidenced by the two lines-of-sight with
similar optical depth ratios.  However, for more detailed studies of density
variations, EVLA observations can and should be employed.

The single largest uncertainty is related to the mean properties of the GMC.
While we have accounted for these uncertainties by adopting a very conservative
range of values for the mean density (covering two orders of magnitude), it is
not entirely clear how the mean density of the cloud should be computed for
comparison to simulations and the analytic distributions.  Since this is a
foreground cloud lying in front of a rich portion of the galactic plane, the
best mass estimates will always come from molecular line observations,
and therefore they are unlikely to be improved unless new wide-field CO
observations are taken, e.g. with CCAT.  

To validate our cloud mean density measurements, we have performed a dendrogram
analysis \citep{Rosolowsky2008c} on the integrated \thirteenco map of the
\GRSMC cloud.  We perform the analysis both on the large-scale $r\sim20$ pc
cloud, tracking down to 10 pc scales, and then on the individual clump that is
directly in front of W49.  We show the cloud density, computed using the
assumptions stated in the text to convert \thirteenco luminosity to mass, for
three different geometrical assumptions described in the caption of Figure
\ref{fig:dendro}.  While the clump densities are potentially higher than we
assumed in the analysis, they are probably not the appropriate numbers to
compare to the simulations we have cited, which are generally simulating entire
molecular clouds and measuring the density distribution within a large box.

\Figure{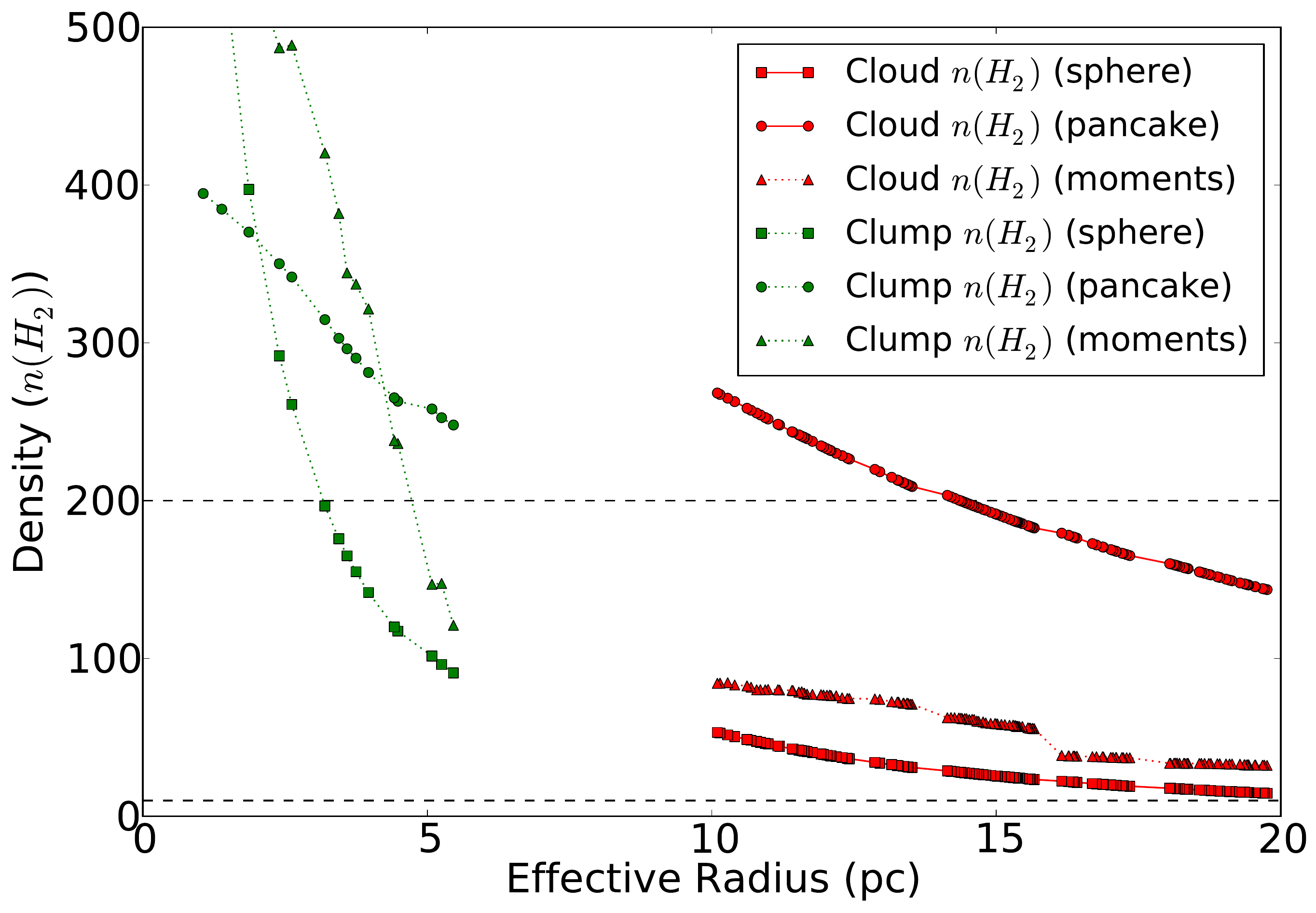}
{Results of a dendrogram analysis of the \GRSMC cloud and the
northernmost \thirteenco ``clump'' within that cloud.  The data points represent
successively higher (and therefore smaller) contours within the integrated \thirteenco
map.  The shapes represent three different methods
for extracting the volume: squares assume spherical symmetry using the
effective radius of the contour, which is proportional to the square root of
the number of pixels.  Circles do the same, but assume that the line-of-sight
radius is 2 pc (i.e., smaller than the observed plane-of-the-sky
dimensions).  The triangles show the volume of an ellipsoid using the moments of the
contoured pixels, with volume $V=4/3 \pi R_{maj}R_{min}^2$.  The black dashed lines
indicate the range of densities allowed in our fits.}
{fig:dendro}{0.5}{0}

\end{document}